\documentclass[conference]{IEEEtran}
\usepackage{comment}
\usepackage{multirow}
\usepackage[dvipsnames]{xcolor}
\usepackage[utf8]{inputenc}
\usepackage{booktabs}
\usepackage{comment}

\ifCLASSINFOpdf
  \usepackage[pdftex]{graphicx}
\else
\fi
\usepackage{hyperref} 

\begin{document}
\title{Assessing the Maturity of Digital Twinning Solutions for Ports}

% Added by me
%\begin{comment}
\author{\IEEEauthorblockN{Robert Klar\IEEEauthorrefmark{1}\IEEEauthorrefmark{2},
Anna Fredriksson\IEEEauthorrefmark{1},
Vangelis Angelakis\IEEEauthorrefmark{1}
}
\IEEEauthorblockA{\IEEEauthorrefmark{1}Department of Science and Technology, Linköping University,
Campus Norrköping, 60 174, Sweden}
\IEEEauthorblockA{\IEEEauthorrefmark{2}Swedish National Road and Transport Research Institute (VTI), SE-581 95 Linköping, Sweden}
E-mail: robert.klar@liu.se, anna.fredriksson@liu.se, vangelis.angelakis@liu.se
}
\maketitle

\begin{abstract} 
Ports are striving for innovative technological solutions to cope with the increasing growth in demand of goods transport, while at the same time improving their environmental footprint. An emerging technology that has the potential to substantially increase the effectiveness of the multifaceted and interconnected port processes is that of  digital twins. Innovation-leading ports recognizing the potential of twinning have already started working on it. However, since there is no clear consensus on what a digital twin of a complex system comprises and how it should be designed, deployed digital twin solutions for ports often differ significantly. This article addresses this issue by initially identifying three core aspect underpinning digital twins of complex systems, such as  ports, and outlining five successive maturity levels based on these aspects' instantiation. These identified aspects and the derived maturity levels are then used to examine real-world cases by critically evaluating existing digital twinning solutions in the port of Singapore, the Mawan port of Shanghai, and that of Rotterdam. These being three of the world's innovation-leading ports, we naturally find in them most of the identified core aspects to be in line with their twinning implementation, which has reached, in all three, a higher level of maturity. Although, our work on maturity levels and core aspects can provide a guideline for designing and benchmarking future digital twinning solutions for ports, the capacity for innovation via twinning, even in the port domain, is highly contextual with key paragon being the availability of financial and technical resources.

\end{abstract}


% \fi
\IEEEpeerreviewmaketitle

\section{Introduction}

%% \subsection{Motivation -or: Why are Seaports of interest?}

\IEEEPARstart{S}{eaports} are striving for innovative technological solutions to cope with the steady growth of transport of goods, in the context of globalization, while complying with new regulations intended to enhance the sustainability of port operations \cite{notteboom2021port}. 
As a result, the scale and complexity of port operations is increasing, requiring more sophisticated and accurate computational models to derive precise planning to meet the demands of the future \cite{chenhao_digital_2018}.
This dynamic and competitive character of the maritime and port landscapes drives the application of new technologies and innovation to enhance performance and increase cooperation and transparency, and attract new business \cite{mudronja_seaports_2020}. 
European ports are focusing on safety, efficiency and sustainability to fulfill a 32 \% increase in energy efficiency according to the 2030 climate and energy framework.
%%European ports are focusing on safety, efficiency and sustainability to comply with the requirements of the 2030 Climate and Energy Framework, with a target of improving energy efficiency by at least 32.5 \% 
\cite{sdoukopoulos2019energy}. In doing this, many ports have launched efforts to give port staff a complete and up-to-date overview of port activities via digital twinning\cite{keegan_driving_2019}. Twinning in this context aims to  enhance ports real-time situational awareness for static, moving, human-controlled or autonomous entities and artifacts, by bringing together geographic, sensor and real-time information. However, the implementation of the digital twin in complex systems such as  ports is still in its infancy, caused in part by the fact that there is no clear consensus in the various sectors of what a digital twin is and how it should be designed. As a result, various leading ports are designing digital twin solutions that differ significantly in terms of their functional scope, and thus their level of maturity. Subsequently, there is a need for tools to benchmark the progress of digital twin implementation between different ports. 

\subsection{Digital Twins and Ports}

Originally developed for supporting manufacturing, digital twinning has attracted a great deal of attention in industry and practice. Moreover, digital twins (DTs), enabling simulation of systems' behaviour in digital form has been referred to as ``a quantum leap in discovering and understanding emergent behavior'' \cite{grieves_vickers_2017}. The potential of digital twins to optimise port processes towards reducing costs or avoiding CO2 emissions, for example, has also been recognised at innovation leading ports \cite{keegan_driving_2019,dalaklis2022port,de2020industry}. It is therefore not surprising that there are already existing twinning solutions catering to the needs of different port stakeholders, twinning different port elements. 
In paper \cite{zhou2022digital}, Zhou et al. implement a digital twin port crane framework based on multi-sensor data, which is able to reproduce the historical crane operation process, simulate the control program, simulate the synchronous mapping and take remote control. Zhou et al. further propose in paper \cite{zhou_analytics_2021} a decision support system with a digital twin-based resilience analysis that assesses a port's resilience to potential disruptive events, taking into account its design, operations and possible predefined post-event recovery measures to mitigate the impact of the disruption. Further decision support related digital twin studies emphasize on integrated crane maintenance under operation in container terminals \cite{szpytko_digital_2021} or on dispatching assistance in port logistics based on a performance forecast \cite{hofmann_implementation_2019}. In paper \cite{li_automated_2021}, Li et al. propose a safety operation optimization framework integrating the digital twin with the AdaBoost algorithm to increase container terminal efficiency and safety. In paper \cite{wang_multi-aspect_2021}, Wang et al. present core techniques for a systematic framework of a digital twin-based model focusing on transport and operations for smart port management.

%%Although these research articles have significantly contributed to the understanding of the potential of digital twinning in ports, there is still a lack of understanding of what %%requirements need to be met in order to achieve a holistic digital twin of the port, leading to different perceptions of what constitutes a digital twin for the port.

\subsection{The (Im)maturity of Digital Twins in Ports}

Although the concept of the digital twin has largely evolved since its coining in 2002 \cite{grieves_vickers_2017}, there is still a lack of standardization, methodologies, and tools for the development and implementation of digital twins \cite{perno2022implementation}. Furthermore, the concept and content of digital twins do not have a precise, uniform definition or even description \cite{zheng2019application}. It is therefore not surprising that there are also problems with the uniform implementation of the digital twin in ports. An additional obstacle is that different port actors usually keep a wide array of, practically, vertical information systems (i.e. with limited or no actual interconnection between them), due to the large number of actors in the port processes and the isolated procurement digitization projects that provided these systems.
Due to this lack of data exchange individual operators are seldom able to efficiently plan the use of resources (short and long term), since it can be  difficult to accurately predict when some of them will be in need, for example, when a port a call will occur \cite{lind2015port}. Finally,  the wide diversity of ports in terms of size, geographical characteristics, governance and institutional frameworks, port functions and port specializations \cite{notteboom2021port}, is creating an additional challenge for a one-size-fits-all digital twinning approach in ports. 

% \newpage


%\subsection{Subsection Heading Here}
%Subsection text here.

%\IEEEpubidadjcol

%\subsubsection{Subsubsection Heading Here}
%Subsubsection text here.

\section{Digital Twins} \label{Digital_Twins}

\subsection{Definition and characteristics of Digital Twins} \label{sec:DT_definition}
Twinning of physical assets, or processes represents a step in the process of digitization and, as such, has been evolving together with technologies supporting its realization (e.g., sensor technology, Internet of things (IoT), cloud computing, big data analytics, and artificial intelligence (AI)) in the last twenty years \cite{qi2021enabling}. Digital twins are  considered to be the pillar of Industry 4.0 and the innovation backbone of the future as they bridge the virtual cyberspace with physical entities \cite{jiang_industrial_2021}. Several definitions of Digital Twins (DT) consider it a virtual representation (replica) of an actual system (AS), which can continuously update with real-time data throughout its lifecycle and can interact with and influence the AS \cite{eramo_conceptualizing_2022}. However, since DT is the subject of study in several disciplines and a tool applied across different disciplines, where practitioners  understand it from their unique professional perspective, there is no uniform definition. Comparing different definitions, it is evident that the more recent definitions of digital twin focus on dynamics, learning, and evolution, rather than just being digital shadows of static objects in the real world \cite{mylonas_digital_2021}. Drawing on the work of Jiang et al. \cite{jiang_industrial_2021}, some of the core aspects of the digital twin are summarised here. \\

\noindent\textbf{(1) Components:} A digital twin and its physical counterpart consist of Physical Entities, Virtual Models, Physical-Digital Connections, Data, and Services, both in the real world, which need to be modelled, and those in the cyber world. 

\noindent\textbf{(2) Temporal span}: A digital twin is designed to mirror its physical counterpart throughout its life-cycle (design, prototyping, manufacturing, deployment, maintenance, and disposal). We must note that each of these phases, comprising different functionalities, may require different timescales at which components will have to operate.  

\noindent\textbf{(3) Functional scope}
    \begin{itemize}
         \item \textbf{Modelling:} A digital twin is a grouping of models and algorithmic components that jointly describe a complex system and allow to estimate the impact of likely outcomes, e.g., to test what-if scenarios and enable predictive maintenance. 
         \item \textbf{Visualization:} A digital twin enables a digital replica of all static and dynamic processes as well as the components of its physical counterpart. 
          \item \textbf{Interaction:} A digital twin is characterized by its bidirectional character, enabling it to directly influence the actual system based on its actions, changes, and predictions. 
        \item \textbf{Synchronization:} The digital twin is continuously updated in a timely manner by various components and processes of the actual system whenever needed. 
        \item \textbf{Self-Improving:} A digital twin is a (self-)improving system that can be progressively improved and extended through the increasing accumulation of data and knowledge over time. 
    \end{itemize}

\begin{table*}[h]
\renewcommand{\arraystretch}{1.3}
    \caption{Comparison of Digital Twin Definitions
    }
    \label{crouch}
    \begin{tabular}{  l  p{12cm}  p{3.4cm} }
        \toprule
\textbf{Domain}      
& \textbf{Definition and purpose of digital twins across different domains}   
& \textbf{Key characteristics} \\\midrule
\multirow{2}{*}{Manufacturing} & \textbf{Definition:} "A digital twin is an integrated multi-physics and multi-scale simulation of a product/system that can model the mechanical, electrical, software, and other discipline-specific properties across its lifecycle" \cite{leng_digital_2021}. & Simulation, Modelling \\
          & \textbf{Purpose:} "Through high-fidelity modeling, real-time interaction and data fusion, DT can reproduce a physical asset or process accurately in the digital world and enable more effective monitoring, optimization, and prediction of the physical counterpart throughout its lifecycle" \cite{tao_digital_2021}.    & Fault detection, Mirroring, Monitoring, Optimization \\
          \hline
\multirow{2}{*}{Smart City} & \textbf{Definition:} "An urban digital twin can be best characterized as a container for models, data, and simulations" \cite{dembski_urban_2020}. & Complexity capture   \\ 
          & \textbf{Purpose:} "The city digital twin is anticipated to construct a link with the real city or the physical counterpart to enhance the visibility of the city and the understanding and analysis of the city’s events and operations. For that purpose, the city digital twin is perceived as enabling technology to promote situational awareness for city management and to provide a city information model; that is, the city digital twin can collect, monitor, and manage city data" \cite{shahat_city_2021}.    & Situational awareness, Monitoring, Analysis, Understanding  \\
          \hline
\multirow{2}{*}{Supply Chain}  & \textbf{Definition:} "A digital SC twin is a model that represents the network state for any given moment in time and allows for complete end-to-end SC visibility to improve resilience and test contingency plans" \cite{ivanov_digital_2021}. & Resilience, Replication, Simulation \\ 
          & \textbf{Purpose:} ”A digital supply chain twin acts as a tool for decision-makers in logistics and supply chain management to holistically improve logistics performance along the whole customer order process through data-driven decision-making" \cite{busse_towards_2021}. & Decision making, Holistic improvements \\
    \hline
    \multirow{2}{*}{Ports}  & \textbf{Definition:} ``A digital twin of a port is a grouping of models and algorithmic components that jointly describe the complex interplay of port processes and operations allowing the characterization, estimation, and prediction of the most efficient operations at the process level, but also for the port as a whole'' \cite{klar_portDT}. & Process optimization, Holistic improvements\\ 
    & \textbf{Purpose:} ``Through inputs from real-time sensors and experience from historical data, a user can identify patterns that led to inefficiencies in the past, get a complete view of current operating conditions, and predict future conditions by simulating what-if scenarios'' \cite{klar_portDT}.    & Simulation, Modelling, Fault detection, Monitoring\\ 
        \bottomrule
    \end{tabular}
    \label{DT_definition}
\end{table*}

A comparison of different definitions of digital twins as illustrated in table \ref{DT_definition}, where aside from the original application of DTs in manufacturing, we bring in definitions from the domains of smart cities and supply chains, since in \cite{klar_portDT} we discuss how ports can be viewed from the infrastructure perspective of the smart city and their relevance to supply chains. Table \ref{DT_definition} illustrates how different application domains have different requirements, particularly in terms of the scale, frequency of updates and predictive capabilities. This distinction is underlined by Mylonas et al. \cite{mylonas_digital_2021} who point out that scale is one of the fundamental differences between DTs in smart manufacturing and those in smart cities, as smart cities are essentially systems of systems and the complexity and heterogeneity of DTs at the urban scale may be orders of magnitude greater than their industrial counterparts. Consequently, a domain-based definition of digital twins might be more advantageous than a general definition that cannot apply to all domains. Although definitions and characteristics differ, table \ref{DT_definition} demonstrates that digital twins among various domains have common objectives, including fault detection, product/process optimization, evaluation of potential operation szenarios, monitoring, and the ambition to save costs while enhancing safety and sustainability.  
\\

\subsection{Different stages of maturity of Digital Twins}
Based on the comprehensive evaluation of existing digital twin solutions by Botín-Sanabria et al. \cite{botin-sanabria_digital_2022}, most DT concepts are still at initial stages (maturity levels 0 to 3 in table \ref{DT_maturity}), and few have started integrating real-time data streams, because capturing, filtering, and processing data in real time is a major challenge, and device malfunctions and poor calibration can lead to anomalies or missing data points. In their mature version, digital twins are more than just Building Information Modeling (BIM) or a 3D model. They can then serve as a data resource that enhances the design of new facilities and the understanding of the condition of existing facilities, verify the as-built condition, perform "what-if" simulations and scenarios, or provide a digital snapshot for future works \cite{dtmaturity}. Consequently, a fully developed DT is expected to have elements of self-adaptiveness in combination with machine learning, simulation, and data processing to enable accurate prediction of specific properties related to performance \cite{eramo_conceptualizing_2022}.

\begin{table*}[ht!]
        \centering
        \renewcommand{\arraystretch}{1.3}
        \caption{Maturity levels for digital twins, adapted from \cite{dtmaturity}}
        \begin{tabular}{p{1cm}p{2cm}p{5cm}p{5cm}} 
            \hline
            Level & State & Requirement & Enabled potential\\
            \hline
            1 & Replication of assets & Digitization of physical assets and their state at moment of capture  (e.g. 2D maps or 3D models) & Awareness of assets, rudimentary decision support \\
            \hline
            2 & Connection & Connect processes and models to static data and metadata of level 1 & Realistic simulations and asset planning \\
            \hline
            3 & Synchronization & Enrich with timely data (sensors and other IoT technologies) & Real-time situational awareness and immersive environments\\
            \hline
            4 & Interaction & Two-way data communication and interaction & Remote control of physical assets and processes\\
            \hline
            5 & Automation & Transparent explainable systems with broad control potential & Autonomous operations optimization and self-maintenance  \\
            \hline
        \end{tabular}
        \label{DT_maturity}
\end{table*}
\section{Ports Digital Twin}\label{port_twin}

\subsubsection{The port context} \label{situational_awareness}
Ports, physically located at the outskirts of (smart) cities, perform a variety of functions, as nodes in transportation chains and hubs of economic activities related to the handling of ships and cargo in the port \cite{nijdam2017port}. 
In addition to the port's primary role as a global hub with the goal of establishing excellent port operations that enable the seamless transfer of goods between the maritime and hinterland networks, ports also serves as industrial clusters as well as an information hubs \cite{zuidwijk2017ports}. 
As ports handle a multitude of processes performed by a variety of actors in parallel, it is increasingly important to improve the overall view of port processes and identify potential bottlenecks to increase efficiency, safety, and sustainability throughout the port ecosystem \cite{heilig2017port}.
Consequently, the implementation of the port's digital twin should provide a high level of connectivity and visibility to provide situational awareness that best benefits the many transportation chains that pass through the port.
Beyond situational awareness, ports also benefit from higher levels of automation, as terminal automation can directly improve performance indicators such as cost, efficiency, safety, and reliability \cite{notteboom2021port}. Increased terminal automation is also needed to cope with the increasing size of cargo ships and growing freight volumes. According to Vis et al. in paper \cite{vis2017container}, operations take place at five different areas, namely at the berth, quay, transportation, yard, and gate, leading to a large magnitude of interlinked processes that can be classified into strategic and operational problems.
A second requirement for the port's digital twin is therefore to provide data-driven analytics to aid holistic decision-making. 
Furthermore, the port is a hub of numerous processes involving multiple actors and dimensions. Ports not only align the interests of employees, management and shareholders, but also serve with a wide range of stakeholders, including terminal operators, vessel operators, railways, shippers, industry associations, municipalities, and government agencies \cite{ashrafi2020review}.
A third requirement for the port's digital twin is therefore to foster cooperation between the various port stakeholders. 

% ideas
%Predictive maintenance based on digital twins of key assets and their components
%optimized cargo planning
%efficient utilization of ressources. 
%most efficient designs and set ups

%By integrating IT processes and systems, processes and procedures can be simplified (e.g. electronic data exchange, IT integration, joint planning, %supply chain integration and integrated ICT and joint ventures) \cite{kim2014sustainability}.
%Ports have grown beyond their original role as simple transshipment hubs, as they are now able to offer a wide range of services and activities to %support the entire supply chain.  \cite{mangan2008port}.
%\textcolor{red}{DT-driven management for container terminal operation taking into account risk prediction}

\subsubsection{Digital twin maturity levels in the port context}\label{collaboration}

Adapting the maturity levels for built environment digital twins of \cite{dtmaturity}, we collapse their two initial levels of fundamental digitization, assuming that no modern digital twin cannot comprise a sufficient asset digitization of their 1st level. So, the first step in twinning ought to comprise the production and curation of any respectively needed or already existing records (such as databases, 2/3D models, etc.) towards capturing the physical components of the system we twin in a digital form. Attaching models of physical or business processes which can capture the effect of real-world events relevant to the assets and their life-cycle, is required to achieve the 2nd level of DT maturity. Then, sensor networks can collect real-time data on port traffic (sea and land), the ecological environment, and the various processes of port operations, enabling a timely connection and mapping from the physical world to the digital world. Thus,
the integration of both static and dynamic data, enables real-time situational awareness by providing knowledge about both less time-sensitive real-world events (such as weather conditions or a storage yard capacity) and more time-sensitive events (e.g. the current available space in a storage yard, or the extend of traffic disruptions in hinterland traffic due to expected snowfall).
The steps described above reflect levels 1-3 in table \ref{DT_maturity} and the digital twin core aspects visualization and synchronization in chapter \ref{DT_definition}. 
Then on, based on the collected operational data, two-way exchange of information would allow control commands to be issued remotely, supported by digital twin simulations, thus embedding the bi-directional nature of the digital twin into the port. In last maturity level, the digital twin should be able to make decisions regarding operations and maintenance autonomously based on real-time and historical data and their further processing in its models.
The consecutive steps upon described here reflect levels 4-5 in table \ref{DT_maturity} and the digital twin core aspects modelling and synchronization in chapter \ref{sec:DT_definition}.

The port digital twin should further provide a platform for collaborative decision making, potentially even across organizational boundaries, where multiple port stakeholders would be networked, and simulations of what-if scenarios predict the effects of different measures on the actor itself, but also on the port as a whole. In addition, the various port stakeholders could already report known difficulties, such as staff shortages or planned maintenance in the platform, whereupon the digital twin's simulation models would already indicate possible consequences and alert the other stakeholders to the potential impact. In the case of port expansion or the introduction of a potential new policy, the digital twin platform would also provide all relevant stakeholders with insight of consequences, thus supporting joint decision-making. Therefore, the various stakeholders could be actively involved in the decision-making process through suggestions and direct testing of these. Consequently, the port's digital twin would allow full remote control based on input from all stakeholders involved, while operating autonomously with full self-governance with complete oversight by the main operator and full transparency to all stakeholders involved. The holistic port digital twin described here includes all the requirements of the maturity levels in table \ref{DT_maturity} reaching the 5th level. 

\section{Maturity Evaluation of three port DT solutions}

\begin{table*}[ht!]
        \centering
        \renewcommand{\arraystretch}{1.3}
        \caption{Analysis of DT characteristics and maturity evaluation of leading ports}
        \begin{tabular}{p{2cm}p{12.7cm}p{2.3cm}} 
            \hline
            Port & Capabilities of their respective digital twin solutions & Maturity level \\
            \hline
            \multirow{2}[2]{*}{Mawan Smart Port} & Mawan Smart Port's fleet management system can monitor the status of partially unmanned vehicles in real time, provide optimal routing and scheduling, and support mixed operations of multiple unmanned trailers \cite{sun2021smart}. & 3 (Synchronization)  \\
            & By accessing container data in the container service management system, Mawan Smart Port dynamically generates containers in 3D digital twin with the same position, appearance, number and type as their physical counterparts, enabling users to control the storage of containers and find them quickly, improving the efficiency of yard and container service management \cite{yu_digital_2022}. & 3 (Synchronization)  \\
            & Mawan Smart Port can access historical operation data to create a historical operation review and to identify the causes of operational bottlenecks, helping to provide solutions to avoid them in the future. In addition, current operating conditions are improved based on insights from historical operating data \cite{yu_digital_2022}. & 4 (Interaction) \\
            & The port is able to access real-time operational data, enabling real-time operational monitoring that overcomes the shortcomings of video surveillance and improves the efficiency of real-time operational planning \cite{yu_digital_2022, sun2021smart}. & 4 (Interaction)  \\
            \hline
            \multirow{2}[2]{*}{Port of Singapore} &  The port of Singapore provides a convenient and smooth platform for information communication between port actors, port related services, freight forwarders and logistics companies \cite{yen2022smart}. & 3 (Synchronization) \\
            & The Singapore smart port uses a traffic monitoring system based on, among other things, traffic monitoring sensors along the main roads, capable of tracking the movement of a truck in the Singapore port in real time, notifying it when the vehicle approaches the main facilities and giving it instructions on how to proceed \cite{wang_multi-aspect_2021}. & 3 (Synchronization) \\
            &  The digital twin of the Port of Singapore has enabled companies to conduct planning based on simulated data from the past, present and future. Costs could be reduced and productivity increased without the need for physical simulations or testing, saving time and resources \cite{mpa_singapore_2022}. & 4 (Interaction)  \\
            & Leveraging the digital twin of container terminals and integrating advanced simulation-based optimization techniques, Sigapore's port digital twin can aid to port operators to find the optimal resource configuration in terms of the number of quay cranes, yard cranes and vehicles in an efficient and accurate way \cite{li2017capacity, chenhao_digital_2018}. & 4 (Interaction) \\
            \hline
            \multirow{3}[3]{*}{Port of Rotterdam} & Equipped with sensors throughout its docks, the Port of Rotterdam is capable to collect real-time data on the environment and water conditions, including air temperature, wind speed, humidity, turbidity, water salinity, current, levels, tides and currents \cite{singh2022applications}. &  3 (Synchronization) \\
            & The Port of Rotterdam's digital platform termed Portmaster provides precise information, including the arrival and departure times of ships. Quays, berths and other port infrastructure can be digitally mapped in Portmaster and linked to information about accessibility and availability \cite{port_of_rotterdam_digital} . & 3 (Synchronization)\\
            & The digital twin uses IoT sensors to enable advanced intelligence by simulating the physical characteristics of the port so that multiple variables can be changed and effectively tested. As a result, the DT models are able to make accurate real-time predictions of arrivals and departures, reducing waiting times and costs \cite{port_of_rotterdam_digital, dalaklis2022port}. & 4 (Interaction) \\
            & Portmaster's asset planning and monitoring, maintenance and services modules help the port to get the most out of their assets and revenues as infrastructure is better utilized and port operations are handled more efficiently \cite{port_of_rotterdam_digital}. & 4 (Interaction) \\
            \hline
        \end{tabular}
        \label{DT_maturity_evaluation}
\end{table*}

The maturity levels of table \ref{DT_maturity_evaluation} are applied to assess the core capabilities of the digital twin in three selected innovation-leading  ports. Singapore's smart port has been known as one of the most cutting-edge technology ports in the world \cite{zhou2021analysis}. The Mawan smart port in the Shenzen province of the PRC, has impressively demonstrated how to upgrade a traditional bulk cargo terminal transforming it into an automated terminal \cite{yu_digital_2022}. Europe's largest commercial port and labeled as the ``world's information port'', Rotterdam is also one of the world's leading ports in the research and application of smart technologies. 

The evaluation of the core functions of the digital twins of the respective ports we present in table \ref{DT_maturity_evaluation} shows that they have progressed up to level 4 . All three ports are equipped with connected sensors, allowing, in real time, both the monitoring of certain parameters (truck movements, water and environment conditions), and port process optimization. Moreover following the definitions we presented in Section \ref{sec:DT_definition}, these ports also use simulation tools to predict the optimal resource configuration, as well as the impact of delays on subsequent terminal processes. Consequently, existing digital twins solutions in leading ports provide the digital twin core functional aspects of modelling, visualization (and monitoring), as well as synchronization and interaction to a certain extend. They further provide situational awareness, analytic capabilities for smart decision making and a a communication platform as identified as the core requirements in section \ref{port_twin}. However, there is still potential for further progress in the functional aspects of interaction and self-improvement, as the digital twin should ideally be able to take decisions autonomously, assessing the impact of each process not in isolation but as part of a number of interconnected processes on the port as a whole. 

\section{Discussion and Conclusion}
The article critically evaluated core aspects of digital twins taking into account their relevance for ports. Based on this evaluation, an existing digital twin maturity assessment from the construction industry was refined and contextually transferred to the port context, without losing its generality. We found that leading ports have already achieved considerable success in implementing digital twins meeting the core aspects presented in section \ref{sec:DT_definition}. This is also confirmed by our assessment of their core functionality using the maturity levels in table \ref{DT_maturity}, where these reach up to level 4 of 5. However, we note that despite considerable success in visualization, real-time data acquisition, modeling and simulation, there is still potential for networked process optimization, ideally performed autonomously and with full transparency in the future. In addition, future digital twin solutions have the potential to be self-learning, i.e., to draw conclusions from the consequences of autonomously made decisions and thereby improve themselves. Even though the three ports presented here already have sophisticated digital twin solutions, it must be kept in mind that this does not reflect the situation at large, as many ports differ significantly from the three presented ports in terms of throughput, and thus also financial resources. Finally, our exploration has epidermically addressed the issue of governance and the aspects of cross-stakeholder communication, needed for- and enabled by the sucessful implementation of DT and the impact of port digitization. Thus, the takeaway is that the core aspects and maturity levels of digital twinning in this study can serve as guidance to other ports and as a tool to technologically evaluate current and future digital twinning approaches.


\ifCLASSOPTIONcaptionsoff
  \newpage
\fi



\bibliographystyle{IEEEtran}
\bibliography{bibtex/bib/references.bib}
%\begin{thebibliography}{1}


%\end{thebibliography}


\newpage
\begin{IEEEbiography}{Robert Klar}
Biography text here.
\end{IEEEbiography}

\begin{IEEEbiography}{Vangelis Angelakis}
Biography text here.
\end{IEEEbiography}

\begin{IEEEbiography}{Joakim Kalantari}
Biography text here.
\end{IEEEbiography}

\begin{IEEEbiography}{Anna Frediksson}
Biography text here.
\end{IEEEbiography}


%\newpage



%\vfill


\end{document}